\documentstyle[epsfig, psfrag,12pt,eqn]{article}
\slshape
\newcommand{\be}{\begin{equation}}
\newcommand{\ee}{\end{equation}}
\newcommand{\bef}{\begin{displaymath}}
\newcommand{\eef}{\end{displaymath}}
\newcommand{\bes}{\begin{eqnarray}}
\newcommand{\ees}{\end{eqnarray}}
\newcommand{\besf}{\begin{eqnarray*}}
\newcommand{\eesf}{\end{eqnarray*}}

\newcommand{\margen}{\hspace{8mm}}

\begin{document}

\title{Linking a Minimal Theoretical Model of Scale Factor Evolution with 
Observational Universe Information}

\author{M.P.Infante$^{(1),(2)}$\footnote{E-mail: infante@mesunb.obspm.fr}, 
N.S\'anchez$^{(1)}$\footnote{E-mail: Norma.Sanchez@obspm.fr}   \\
{\footnotesize {\it (1) Observatoire de Paris-DEMIRM. 61, Avenue de 
l'Observatoire, 75014 Paris, FRANCE }} \\
{\footnotesize {\it (2) Dpto. F\'{\i}sica Te\'orica, Univ. de Zaragoza. 
Pza.San Francisco, 50009 Zaragoza, SPAIN}}}
\date{\empty}
\maketitle

\begin{abstract}

We study several approaches for constructing 
a minimal model of Universe evolution by
matching different stages of scale factor
laws. We discuss the continuity in the transitions
among the stages and the time variables involved. 
We develop a way to modelize these transitions
without loss of observational predictability. 
The key is the scale factor and its connection with
minimal but well stablished observational information
(transition times and expansion ratii).
This construction is clearly
useful  in metric perturbations type computations,
but it can be applied in whatever
subject dealing with a cosmology involving 
different evolution stages.

{\bf Report numbers:} gr-qc/9907079; DFTUZ/99/12.

\end{abstract}
\pagebreak[4]

\section{Introduction}
\setcounter{equation}0

{\margen} Cosmology has a rich variety of theories giving rise to different
evolution laws for the Universe. Realistic models for the Universe must
attach with the available observational data. Among them, the
age of Universe and some known transition times must be considered.

From the theoretical point of view, the usual process consists in
matching successives stages having different evolution laws each one.
Particular interest is devoted to works computing the generation of
metric perturbations, since they must handle the transitions among
different evolution stages. Some minimal conditions of continuity 
(of scale factor and the perturbations) are required.

Different ways for handling these continuity conditions can
be found (\cite{gk93}-\cite{gv93}). Some of them uses impose 
continuity on the cosmic time 
evolution although discontinuity in conformal time itself 
(\cite{abha},\cite{Sah}), while others make a continuous matching
of the scale factor in conformal time variable 
(\cite{gk93},\cite{A},\cite{buon},\cite{bggv},\cite{gv93}). 
Both cases would have an imprint in
the successive study of the process of interest like metric 
perturbations propagation, since of the parameters involved in the
linking of the different stages from the very early to the
current one. At time of confronting with observations, the 
parameters left by the linking process can play a crucial role.

The situation is still confuse when time scales must be fixed
on those minimal models. Some kind of scale factor normalizations
(\cite{abha}) or energy scales at transitions (\cite{gv93})
are usually imposed in order to overcome the loss of predictability
of the temporal variables used.

Our scope is to review this process in constructing the evolution
of these minimal Universe models. Lacking a complete underlying
theory describing the total Universe evolution or at least, the detail
of transitions between stages, one must join these stages in 
order to compose an {\it asymptotically} suitable description. The way
how handle each stage and how make the matching among them is
the subject of this paper. We try to apply a different and more 
practical treatment which preserves
the level of
{\it predictability} of the minimal models discussed.

From the theoretical point of view, different evolution laws can be 
provided for the Universe: inflationary expansion of power law type, De 
Sitter type, inverse power type, deccelerated and standard expansion stages. 
From the observational side, the Universe appears to be homogeneus
and isotropic at sufficiently large scales, following the distribution of
clusters of galaxies. From the measurements of redshift in the spectra
of galaxies, the Universe has a metric in expansion. 
The measurements of 
current Hubble factor give us an order of the age of Universe of 
${\mathcal{T}}_0 \sim 10^{17} s$, at least as upper limit.
We find ourselves in an
Universe with an expansion rate matter
dominated-type . We know too the previous existence
of a hotter and denser radiation dominated expansion, as can be
assured from the current existence of the Cosmic Microwave
Background Radiation.
The end of radiation
dominated stage is estimated at ${\mathcal{T}}_m \sim 10^{12} s$.

Finally, a mixed theoretical-observational interpretation had led to 
suppose the existence of an inflationary stage in the very early
beginning. This is the so called inflationary paradigm, whose exponential 
version looks able to solve the cosmological problems on closure and
flatness of Universe, large scale homogeneity and baryon number.
(See for example \cite{kb} and \cite{lin}).

To explain the whole Universe evolution would mean to explain 
phenomena at very different energy scales. At the date, we lack of 
such a complete theory. Therefore, the tools of theoretical cosmology 
are theories with their own ranges of validity. The ideal total
description is substituted by a description made with successive
stages, giving a step-by-step evolution. Thus,
we find the need of joining the different and in principle
discontinuous evolution stages in order to make the most
approximative description.

It leds us to deal with modelized transitions, {\it not necessarily
corresponding to real transitions}. In the minimal models,
these modelized transitions are very limited with respect
to describe and account the real transitions. The logical
choice of a continuous time variable is yet not trivial and
different options are possible. Further computations
as gravitational waves and imprints on CMB 
are signed by these transitions and can be
signed by these choices. 

We study this problem by applying a more practical
point of view: closer to the observational Universe, better
description. This guideline has suggested us an appropiate 
treatment of the temporal variables and transitions involved. 

What do we mean by a minimal model? An {\it asymptotically} valuable 
description of the scale factor evolution running on a temporal
variable. Real transitions among the different stages, probably discontinuous, 
are modelized trough this mathematical temporal-type 
variable in a more convenient way. Further linking with observational Universe
information preserves the prediction capability of the descriptive 
minimal model thus constructed.
Every theoretical cosmology requires as its basic tool 
a scale factor and a temporal variable over which runs. The
minimal observational Universe information is translated on the
theoretical model evolution through such scale factor function $a(t)$
and its transitions (\cite{gksqz}).
Suitable temporal variables for this process of translation are
not a trivial subject. A compromise among the theoretical computational 
ability and the observational Universe information contained is required.

We find the particular proper times in each stage not to be
appropiated variables in the matching process,
because their inherent discontinuity in a minimal model. 
Computations make desiderable to have a minimal conditions
of continuity of scale factor and temporal variable. 
One can ask if real transitions are continuous 
at the level of such minimal descriptions. The answer is 
probably "not". Now, it makes necessary to reinterpretate
the meaning and role of the temporal variable involved in this
minimal model descriptions and distinguish it from the
physical proper time.

Need of deal with different temporal variables and the 
transformations
relating them can be seen at least as the prize to pay for our 
lack of knowledge 
about the dynamics of real transitions among the cosmological
different evolution stages. With this lacking of
knowledge it would be impossible to construct a continuous, full
model for the scale factor including the transitions. We overcome
this problem by defining auxiliary {\it descriptive} 
temporal variables. These variables make a closer mapping of the proper cosmic
time. The freedom in their boundary conditions can
be used in order to make them satisfactorily continuous
through transitions. With the use of such an intermediate tool we must
rewrite the evolution laws in these variables.

A careful handling allows us to maintain a link with the observational
temporal parameters and thus, abilitates our construction to be
worked out in order to further extraction of other observational
cosmological phenomena. The minimal observational Universe information 
introduced are the time scales, by taking the well stablished standard
values for beginning of radiation dominated stage (${\mathcal{T}}_r
\sim 10^{-32} s$), the beginning of matter dominated stage 
(${\mathcal{T}}_m \sim 10^{12} s$) and the proper scale factor
expansion ratii to be reached in each one of the stages.

In the next, we develop the descriptive cosmic time-type variables
for minimal models with an inverse power inflationary stage. We have 
chosen this kind of inflationary behaviours since they appear
typically in the context of String Cosmology. 
The simple guidelines here showed
could be easily followed with whatever other inflationary
type behaviour.  The scale factor description obtained reaches
the proper scale factor ratii in each stage. The transitions
among different stages are modelized as sudden, continuous and smooth.
All parameters and variables in the description have been linked with
observational transition times and proper cosmic time evolution.

We analyze also the conformal
time variable and the corresponding scale factor. 
This construction is
made over descriptive temporal variables. We elaborate
transformation rules in order to keep the link among the conformal
time variable and observational times. Conditions of continuity
and simplicity can be imposed on the most general construction
of conformal time variables.
The scale factor evolution
thus obtained has sudden, continuous and smooth transitions both in the 
cosmic time description and in the conformal time one. This feature of
our treatement is not an usual property for most cosmological minimal models.

We elaborate also another conformal time construction. It is linked to 
observational information and proper cosmic times in a more direct
way. The scale factor in this variable can be written with sudden
and instantaneous transitions, but they are not smooth.

In appendixes we report some other related aspects. We apply the
descriptive variables treatment to a minimal model with arbitrary
power law inflationary stage. The possible modelization of
transitions as brief intermediate stages without descriptive variables
is considered also. Finally, an alternative construction of descriptive
variables is included.
\section{The Temporal Coordinate}
\setcounter{equation}0

\margen Solutions  describing an inflationary stage $(I)$, radiation
dominated stage $(II)$ and matter dominated stage $(III)$ can be obtained 
from different early Universe theories providing cosmological backgrounds. 
We consider them extracted in
the proper cosmic time gauge, where the metric is written as:
\bef
ds^2 = d t^2 - a(t)^2  dx^2
\eef
$a(t)$ is the scale factor in cosmic time variable, giving the 
expansion rate of the metric. Notice here 
$t = c {\mathcal{T}}$. (In natural units, where $\hbar = G
= c = 1$, the equality among variables $t$ and ${\mathcal{T}}$ holds).

 In order to construct a minimal model that describes the
evolution of the scale factor, we will consider these successive three 
stages in descriptive cosmic time type variables as having 
instantaneous and continuous transitions at times $\bar{t_1}$ and 
$\bar{t_2}$.  
With full free parameters, we write our model as:
\bes \label{descr}
\bar{\bar{a_I}}(\bar{\bar{t}}) & = & \bar{\bar{A_{I}}}
{(\bar{\bar{t_I}}-\bar{\bar{t}})}^{-Q} \ \  \ \ \ \ \ \ \ {\bar{\bar{t}}} 
< \bar{t_1}\\
\bar{a_{II}}(\bar{t}) & = & \bar{A_{II}} {(\bar{t}-\bar
{t_{II}})}^R  \nonumber \ \ \ \ \ \ \  {\bar{t_1}} < {\bar{t}} < {\bar{t_2}} \\
a_{III}(t) & = & A_{III} {(t - t_{III})}^M  \ \ \ \ \ \ \ \  {\bar{t_2}} < t
\nonumber
\ees
Here the first stage $(I)$ describes an accelerated expansion provided
${\bar{\bar{t}}}_I>{\bar{t}}_1$ and $Q>0$. This is the inverse power
inflationary stage. In appendixes we consider the case for arbitrary power law
inflationary stages. The behaviours of radiation and matter dominated
stages are identified by power law dependences with exponents
$R=\frac{1}{2}$ and $M=\frac{2}{3}$ in cosmic time gauge.
By imposing the continuous matching of the scale factor and 
its first derivative at the transition times $\bar{t_1}$ and $\bar{t_2}$, 
give us four 
constraints on the six free parameters. We have choose to express the 
parameters ${\bar{\bar{t_I}}}$, ${\bar{\bar{A_I}}}$, $t_{III}$, $A_{III}$ 
in function of ${\bar{t_{II}}}$, ${\bar{A_{II}}}$. We have:
\bes 
\bar{\bar{t_I}} & = & \bar{t_1} \left(1+{Q \over{R}}\right) - \bar{t_{II}} 
{Q \over{R}} \label{match1}\\
\bar{\bar{A_I}} & = & \bar{A_{II}} {\left({Q \over{R}}\right)}^Q {(\bar{t_1} - 
\bar{t_{II}})}^{R+Q} \nonumber \\
t_{III} & = & \left(1 - {M \over{R}}\right) \bar{t_2} + {M \over{R}} 
\bar{t_{II}} \nonumber \\
A_{III} & = & \bar{A_{II}} {\left({R \over{M}}\right)}^{M} {(\bar{t_2} - 
\bar{t_{II}})}^{R-M} \nonumber 
\ees
The inflationary stage is described with a time variable 
$\bar{\bar{t}}$ and the radiation stage with a variable ${\bar{t}}$.
Notice that these cosmic time-type variables (and $t$ of
the matter dominated stage) are not at priori exactly equal to 
the physical proper time coordinate (multiplied by $c$ in no natural
units), but transformations (dilatation plus translation) of it.

We will find relations between these variables and the physical time coordinate
(multiplied by $c$, $t=c {\mathcal{T}}$) which allow to put in contact
this description with the minimal and well stablished observational
Universe information: the scale times (standard transition times) and 
the proper scale factor expansion ratii.
We understand that a descriptive temporal 
variable can provide a better scale factor description.  
The cosmic time type variables here introduced act as intermediate
vehicles in the process of linking observational Universe and
theoretical evolution model. The minimal information about the former
one is imprinted on transformations giving raise such descriptive
variables, while remaining parameters can be fixed in order 
to have a suitable scale factor continuity for cosmology purposes. 

\section{Link with the Observational Universe}
\setcounter{equation}0

\subsection{The Observational Universe}

\margen We consider now some properties of the observational
Universe evolution. Taking the simplest option, we consider the ``real''
scale factor evolution to have transitions at $t_r$
(beginning of radiation dominated stage) and $t_m$ (beginning  of matter
dominated stage), given by the standard observational values.
 We define a beginning 
of inflation at $t_i$ and corresponding current 
time $t_0$. Equality in the temporal dependence exponents in the 
real and descriptive scale factors is preserved, the other free
parameters are related by transitions.
\pagebreak[3]

In order to have the right temporal dependence (following the real duration
and scale factor expansion of the observational Universe) for a radiation 
dominated behaviour and a matter dominated behaviour, we must consider there
a scale factor without additive constants in their proper cosmic time 
dependences. 
Notice as a consequence, we can not consider continuity in first 
derivative of scale factor in the ``real'' transitions. Continuity on
the scale factor itself is supposed by considering real transitions
enoughly briefs and smooths for not requiring be described as 
intermediate stages.
\bes \label{real}
a_I(t) & = & A_I {(t_I - t)}^{-Q} \ \ \ \   t \in (t_i, t_r)  \\
a_{II}(t) & = & A_{II} \: t^R \ \ \ \ \ \ \ \ \  t \in (t_r, t_m) \nonumber \\
a_{III}(t) & = & A_{III} \: t^M \ \ \ \ \ \ \ \ \ t \in (t_m, t_0) \nonumber
\ees

Without knowing the dynamics of the real transitions among stages, 
this approach still allow us to extract a robust constraint from
the observational knowledge of our Universe: the
scale factor expansion (or scale factor ratii) reached in each one of
the three stages considered. Meanwhile the duration of a stage,
the scale factor will expand, in the inflationary stage:
\be \label{rin}
{{a_{I}(t_r)}\over{a_{I}(t_i)}} = {\left({{t_I-t_r}\over{t_I-t_i}}\right)}
^{-Q} \ \ \ \ \ ,
\ee
in the radiation dominated stage:
\be \label{rrad}
{{a_{II}(t_m)}\over{a_{II}(t_r)}} = {\left({{t_m}\over{t_r}}\right)}^R
\ \ \ \ \ ,
\ee
and in the matter dominated stage:
\be \label{rmat}
{{a_{III}(t_0)}\over{a_{III}(t_m)}} = {\left({{t_0}\over{t_m}}\right)}^M
\ee

We consider the standard observational values for cosmological times.
The radiation-matter transition held at: 
\bef
{\mathcal{T}}_m \sim 10^{12} s \ \ \ , 
\eef
the beginning of radiation stage at
\bef
{\mathcal{T}}_r \sim 10^{-32} s \ \ \ ,
\eef
and the current age of the Universe
\bef
{\mathcal{T}}_0 \sim {H_0}^{-1} \sim 10^{17} s \ \ \ \ \ .
\eef
The exact numerical value of ${\mathcal{T}}_0$ turns out not be crucial here.

\subsection{Linking Model and Observations}

\margen We can impose now to our description to satisfy stage by stage the 
same right scale factor ratii as the observable evolution.
For the matter dominated stage, it is convenient to have its
descriptive scale factor written in proper cosmic time, since it is
within this stage where observational information must be extracted and
provided.
Then, we fix one of the two still free parameters in matching 
eqs.(\ref{match1}) by eliminating 
in eq.(\ref{descr}) the additive constant $t_{III}$:
\be \label{tiii}
t_{III} \equiv 0
\ee

   Now, in order to have the right scale factor ratio during the whole
matter dominated stage, the beginning $\bar{t_2}$ must coincide exactly
with the beginning of matter dominated stage $t_m$.
\be \label{t2}
\bar{t_2} \equiv t_m
\ee

This first link with observation translates immediately on the 
radiation dominated stage description. 
From eq.(\ref{match1}) and with the obtained conditions 
eqs.(\ref{tiii}) and (\ref{t2}), the additive constant $t_{II}$ is fixed.
\be \label{tii}
\bar{t_{II}} = \left(1 - {R \over{M}}\right) t_m
\ee

In order to fix the beginning of radiation stage $\bar{t_1}$, we use the 
observational value of scale factor expansion during the whole stage 
eq.(\ref{rrad}) and make it coincident with the value obtained in 
the descriptive scale factor model. By this way:
\bef
{{\bar{a_{II}}(\bar{t_2})}\over{\bar{a_{II}}(\bar{t_1})}} =
{{\bar{A_{II}}{(\bar{t_2} - \bar{t_{II}})}^R}\over{{\bar{A_{II}}} 
{(\bar{t_1}-\bar{t_{II}})}^R}} =
 {\left({{t_m}\over{t_r}}\right)}^R
\eef 
which gives the relation:
\besf  
\bar{t_1} & = & t_r + \left(1 - {{t_r}\over{t_m}}\right) \bar{t_{II}} \ \ \ .
\eesf
With the condition eq.(\ref{tii}), we obtain the
expression of $\bar{t_1}$ as function of the observational transitions 
$t_r$ and $t_m$:
\bes \label{t1} 
\bar{t_1} & = & {R \over{M}} t_r + \left(1 - {R \over{M}}\right) t_m 
\ees

We can interpolate this process for every time $t$ belonging to the radiation 
dominated stage, and require equally to the model to give the corresponding
scale factor expansion ratio at $\bar{t}$. Thus, it is possible to get a 
relation linking the temporal variable ${\bar{t}}$ with the physical
time $t$:
\bes \label{tbar} 
\bar{t} & = & {R \over{M}} t + \left(1 - {R \over{M}}\right) t_m 
\ees

The conditions on $\bar{t_1}$ have fixed yet the end of the descriptive
inflationary stage. For fixing the corresponding beginning at
${\bar{\bar{t_i}}}$, the same above guidelines are followed.
The total expansion during the
inflationary stage (\ref{rin}) must be:
\bef
{{\bar{\bar{a_{I}}}(\bar{t_1})}\over{\bar{\bar{a_{I}}}(\bar{\bar{t_i}})}} =
{{\bar{\bar{A_{I}}}{(\bar{t_1} - \bar{\bar{t_{I}}})}^{-Q}}
\over{{\bar{\bar{A_{I}}}} 
{(\bar{\bar{t_i}}-\bar{\bar{t_{I}}})}^{-Q}}} =
{\left({{{t_r} - {t_{I}}}}\over{{{t_i}-{t_{I}}}}\right)}^{-Q}
\eef 
From this we obtain:
\bef
\bar{\bar{t_i}} = {{{\bar{\bar{t_I}}}\:(t_i-t_r) \: + \: \bar{t_1}\:
(t_I-t_i)}\over
{(t_I - t_r)}}
\eef
and making use of relations (\ref{t1}), (\ref{t2}) and the matching 
equations (\ref{match1})
\be \label{ti}
\bar{\bar{t_i}} = 
 \left({R \over{M}} + {Q\over{M}}{{t_r-t_i}\over{t_r-t_I}} \right)t_r
+ \left(1 - {R \over{M}}\right) t_m
\ee
Again, we can interpolate this relation for all $t$ in the inflationary 
stage and obtain the relation for the temporal variable $\bar{\bar{t}}$ in
terms of the observational transition values $t_r$ and $t_m$:
\be \label{tbarbar}
\bar{\bar{t}} = 
 \left({R \over{M}} + {Q\over{M}}{{t-t_i}\over{t_r-t_I}} \right)t_r
+ \left(1 - {R \over{M}}\right) t_m 
\ee

\begin{figure}
\psfrag{a}{$a_I(t)\sim{\left(t_I-t\right)}^{-Q}$}
\psfrag{b}{$a_{II}(t)\sim{\left(t\right)}^{R}$}
\psfrag{c}{$a_{III}(t)\sim{\left(t\right)}^{M}$}
\psfrag{i}{$\mathbf{{\bar{t_1}}}$}
\psfrag{j}{$\mathbf{{\bar{t_2}}}$}
\psfrag{f}[bl]{${\bar{\bar{a_I}}({\bar{\bar{t}}})}\sim
{\left({\bar{\bar{t}}}_I-
{\bar{\bar{t}}}\right)}^{-Q}$}
\psfrag{g}[B]{${\bar{a_{II}}({\bar{t}})}\sim{\left({\bar{t}}-
{\bar{t}_{II}}\right)}^{R}$}
\psfrag{h}[B]{$a_{III}(t)\sim{\left(t\right)}^{M}$}
\psfrag{d}[t]{$\mathbf{{t_r}}$}
\psfrag{m}[t]{$\mathbf{{t_i}}$}
\psfrag{n}{$\mathbf{\bar{\bar{t_i}}}$}
\psfrag{e}[t]{$\mathbf{{t_m}}$}
\psfrag{A}{{\bf INF.}}
\psfrag{B}{{\bf RAD.}}
\psfrag{C}{{\bf MAT.}}
\includegraphics[width=130mm]{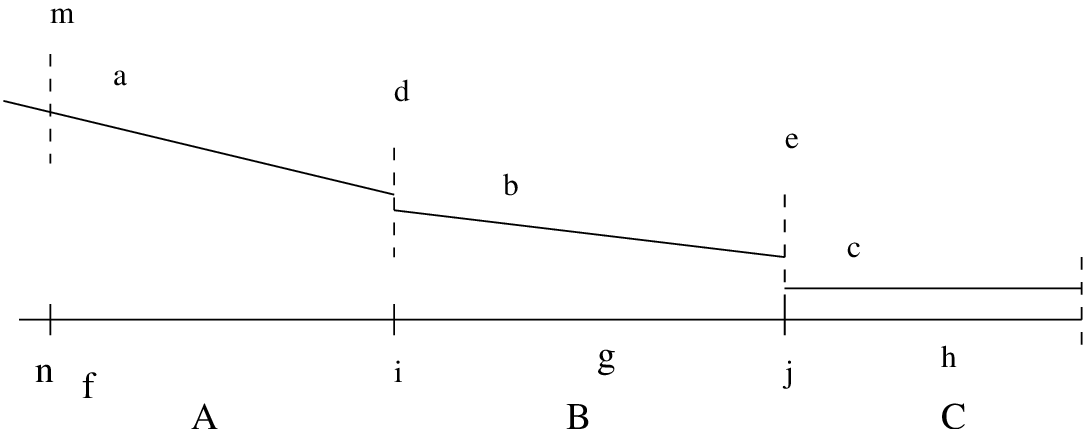}
\caption{Linking with observational information by descriptive
variables}
\vspace{5pt}
{\parbox{140mm}{\footnotesize Representation of the linking process with
observational Universe information. The theoretical minimal model is found 
in the lower side, running on variables ${\bar{\bar{t}}}$, ${\bar{t}}$, $t$ 
with sudden and continuous transitions on such variables at ${\bar{t_1}}$ and
${\bar{t_2}}$. The observational Universe information is represented in
the upper side, whose dynamics at transitions $t_r$ and $t_m$ can be considered
nearly sudden but in any case discontinuous which respect to proper cosmic
times. Expressions relating both pieces are given in the text.}} 
\end{figure}

\subsection{Summary of Scale Factor Description}

\margen Equations (\ref{rin}-\ref{rmat}) have allowed us
to link the model for the scale factor 
eqs.(\ref{descr}-\ref{match1}) with observational Universe
information. 
The scale factor written in cosmic time-type variables
have thus the expression: \enlargethispage{5mm}
\bes \label{Descr}
\bar{\bar{a_I}}(\bar{\bar{t}}) & = & \bar{\bar{A_{I}}}
{(\bar{\bar{t_I}}-\bar{\bar{t}})}^{-Q} \ \ \ \ \ \ \ 
{\bar{\bar{t_i}}}  <  {\bar{\bar{t}}} < {\bar{t_1}}   \\
\bar{a_{II}}(\bar{t}) & = & \bar{A_{II}} {(\bar{t}-\bar{t_{II}})}^R 
 \ \ \ \ \ \ \ \   \bar{t_1} < \bar{t}  < \bar{t_2}\nonumber \\
a_{III}(t) & = & A_{III} {(t)}^M 
\ \ \ \ \ \ \  \ \ \ \ \bar{t_2} <  t < {t_0} \nonumber
\ees
with sudden and continuous transitions at $\bar{t_1}$ and $\bar{t_2}$ for 
both the scale factor 
and first derivatives with respect to the descriptive cosmic time-type 
variables $\bar{\bar{t}}$ eq.(\ref{tbarbar}), $\bar{t}$ eq.(\ref{tbar}) 
and $t$. In such a way, the parameters in function of transitions
and global scale factor $\bar{A}_{II}$:
\bes 
\bar{\bar{t_I}} & = & \bar{t_1} \left(1+{Q \over{R}}\right) - \bar{t_2}
\left({Q \over{R}}-{Q \over{M}}\right) \ \ \ \ \ \  , \ \ \ \ \ \ 
\label{MAtch1}
\bar{t_{II}}  =  \left({1-{R\over{M}}}\right) \bar{t_2}  \\
\bar{\bar{A_I}} & = & \bar{A_{II}} {\left({Q \over{R}}\right)}^Q 
{\left(\bar{t_1} - (1-{R\over{M}})\bar{t_2}\right)}^{R+Q} \ \ \ , \ \ \ 
A_{III}  =  \bar{A_{II}} {\left({R \over{M}}\right)}^{R} {\bar{t_2}}^{R-M} 
\nonumber 
\ees 
We remember that transitions $\bar{t_1}$, $\bar{t_2}$
and the beginning of the inflationary stage description $\bar{\bar{t_i}}$ are
expressed in terms of the standard observational times $t_r$ and $t_m$ as:
\bes  
\bar{t_1} & = & {R \over{M}} t_r + \left(1 - {R \over{M}}\right) t_m 
\ \ \ \ \ \ , \ \ \ \ 
\bar{t_2}  =  t_m \label{Ti} \\
\bar{\bar{t_i}} & = &
 \left({R \over{M}} + {Q\over{M}}{{t_r-t_i}\over{t_r-t_I}} \right)t_r
+ \left(1 - {R \over{M}}\right) t_m \nonumber
\ees
Parameters of scale factor (\ref{Descr}) satisfy also the corresponding
equations (\ref{match1}). With help of 
eqs.(\ref{tiii}),(\ref{tii}) and (\ref{Ti}), these parameters can
be written too in terms of
$t_r$, $t_m$ and the global scale factor $\bar{A_{II}}$:
\bes \label{Match1}
\bar{\bar{t_I}} & = & {t_r} {\left({R\over{M}}+{Q\over{M}}\right)} + {t_m}
{\left({1 - {R \over{M}}}\right)} \ \ \ , \ \ \  
\bar{t_{II}}  =  \left({1-{R\over{M}}}\right) t_m  \\
\bar{\bar{A_I}} & = & \bar{A_{II}} {\left({Q \over{M}}\right)}^Q 
{\left({R \over{M}}\right)}^R {t_r}^{R+Q} \ \ \ \ \ \ \ , \ \ \  \
A_{III}  =  \bar{A_{II}} {\left({R \over{M}}\right)}^{R} {t_m}^{R-M} 
\nonumber \ees

The descriptive scale factor eq.(\ref{Descr}) can be expressed 
also in the proper cosmic time variables as:
\bes \label{Dest}
\bar{\bar{a_I}}(t) & = & \bar{A_{II}} {\left(\frac{R}{M}\right)}^R
{t_r}^{R+Q} \: {\left(\frac{t_I}{t_r}-1\right)}^Q\:{(t_I-t)}^{-Q} \\
\bar{a_{II}}(t) & = & \bar{A_{II}} {\left(\frac{R}{M}\right)}^R \: t^R 
\nonumber \\
a_{III}(t) & = & \bar{A_{II}} {\left(\frac{R}{M}\right)}^R {t_m}^{R-M} \:t^M
\nonumber \ees
but remember that not fully continuity (and not in first derivatives)
helds on proper cosmic time. 
It could be considered sudden instantaneous transitions at
$t_r$ and $t_m$ in such a way that the scale factor is continuous:
\besf
\bar{\bar{a_I}}(t_r) & = & \bar{a_{II}}(t_r) \\
\bar{a_{II}}(t_m) & = & a_{III}(t_m)
\eesf
but there is not continuity in the first time derivative of the scale factor. 
In the inflation-radiation 
dominated transition $t_r$, this continuity condition would be satisfied
by models whose inflation scale factor follows:
\be \label{conti}
t_I = \left( 1 + \frac{Q}{R} \right) t_r
\ee
and is in anycase impossible to obtain derivative continuity at the radiation 
dominated-matter dominated transition $t_m$. 
In sake of generality we will
not consider eq.(\ref{conti}) holds in the following.

By the other hand, in the transformed descriptive variables, the 
transitions are
much more smooth. In fact, we know that 
$\bar{\bar{a_I}}(\bar{t_1}) = \bar{a_{II}}(\bar{t_1})$ and 
$\bar{a_{II}}(\bar{t_2}) = a_{III}(\bar{t_2})$.
It holds also:
\bef
\left. {\frac{d{\bar{\bar{a_I}}}(\bar{\bar{t}})}{d{\bar{\bar{t}}}}}
\right|_{{\bar{t_1}}^-} =  
\left. {\frac{d{\bar{a_{II}}}(\bar{t})}{d{\bar{t}}}}\right|_{{\bar{t_1}}^+}  
\ \ \ \ \ \ \  , \ \ \ \ \ \ \  
\left. {\frac{d{\bar{a_{II}}}(\bar{t})}{d{\bar{t}}}} \right|_{{\bar{t_2}}^-}  
= \left. {\frac{d{a_{III}}(t)}{d{t}}} \right|_{{\bar{t_2}}^+}
\eef
Notice that:
\be \label{tat}
\frac{d\bar{\bar{t}}}{dt} = -\frac{Q}{M} \frac{t_r}{t_r-t_I}
\ \ \ \ \ \ \ , \ \ \ \ \ \ \ \frac{d \bar{t}}{dt} = \frac{R}{M}
\ee
These are the factors absorbing the discontinuity in the derivative with 
respect 
to proper cosmic time and allowing 
continuity in the first derivative of scale
factor in descriptive variables $\bar{\bar{t}}$ and $\bar{t}$ at 
transitions $\bar{t_1}$ and $\bar{t_2}$.

The minimal model thus constructed reaches the same expansion ratii 
in each stage and with the same evolution laws than extracted from
information about observational Universe.
But differently from that, this model has sudden, continuous and smooth 
transitions among the three stages. A feature making possible this effect 
is a change in the
duration of stages. When measured in descriptive variables, matter
dominated stage has the same duration than in observational description.  
Not the same for radiation dominated  and inflationary stages:
\besf
\left.{\Delta \bar{t}}\right|_{RAD} & = & \frac{R}{M} 
\left.{\Delta t}\right|_{RAD}  \\
\left.{\Delta {\bar{\bar{t}}}}\right|_{INF} & = & \frac{Q}{M}
\frac{t_r}{t_I-t_r} \left.{\Delta t}\right|_{INF}
\eesf
For usual values of  $R$ and $M$, the radiation dominated stage suffers
a contraction in descriptive variables. For inflationary stage, it depends
of involved parameters. Contraction takes place in this stage if
$t_I >  \left(\frac{Q}{M}+1 \right)t_r$.

\section{The Conformal Time Coordinate}
\setcounter{equation}0

\margen In the last Section we have obtained a minimal model for 
the scale factor, working with descriptive variables of cosmic time type.
We have  related those with the observational transition times. 
It is also convenient to translate it in terms of
the conformal time.
This variable is defined as $d\eta = \frac{dt}{a(t)}$. The metric
in conformal time gauge is expressed as:
\bef
ds^2 = {a(\eta)}^2 \left(d\eta^2 - dx^2\right)
\eef
Following our definition of cosmic time type variable $t=c \mathcal{T}$,
the variable $\eta$ will have too dimensions of lenght.

\subsection{The general case}

\margen We construct now the conformal time over the descriptive cosmic time
variables, with which the scale factor is satisfactorily continuous.
That means we will have a particular conformal time variable for each stage, 
not only oughted to the different shape of the scale factor but also 
as consequence
of the different temporal variables used in its description. Thus, following
eqs.(\ref{Descr}) we define:
\bes
d \: \bar{\bar{\eta}} & = &  \frac{d{\bar{\bar{t}}}}{\bar{\bar{a_I}}
(\bar{\bar{t}})} 
\ \ \ \ \ {\bar{\bar{t_i}}}  <  {\bar{\bar{t}}} < {\bar{t_1}}  
\label{etain}  \\
d \: \bar{\eta} & = & \frac{d{\bar{t}}}{\bar{a_{II}}(\bar{t})} 
\ \ \ \ \  \bar{t_1} < \bar{t}  < \bar{t_2}  \nonumber \\
d\: \eta & = & \frac{d{t}}{a_{III}(t)}  
\ \ \ \ \bar{t_2} <  t < {t_0}   \nonumber
\ees
We integrate these equations in order to obtain expressions
for the conformal time variable in each stage. 
Let be the parameters $\eta_i$, $\eta_1$ and $\eta_2$ the conformal 
time values corresponding to the beginning of each stage of the
minimal model at $\bar{\bar{t_i}}$, $\bar{t_1}$ and $\bar{t_2}$. 
\bef
\eta_i\: = \: \bar{\bar{\eta}}(\bar{\bar{t_i}}) \ \ \ \ , \ \ \ \ 
\eta_1\: = \: \bar{\eta}(\bar{t_1}) \ \ \ \ , \ \ \ \ 
\eta_2\:  = \: \eta(\bar{t_2}) 
\eef
Now, eq.(\ref{etain}) yields the variable ${\bar{\bar{\eta}}}$ 
for the inflationary stage
\be \label{bbet}
{\bar{\bar{\eta}}} \: = \: \eta_i \: + \: \frac{1}{(Q+1){\bar{\bar{A_{I}}}}} 
\left[{(\bar{\bar{t_I}}-\bar{\bar{t_i}})}^{Q+1}-
{(\bar{\bar{t_I}}-\bar{\bar{t}})}^{Q+1}\right] \ \ \ \ \ \  
\bar{\bar{t_i}}<\bar{\bar{t}}<\bar{t_1}
\ee
For the radiation dominated stage, the corresponding conformal time variable 
$\bar{\eta}$ takes the form
\be \label{bet}
\bar{\eta} \: = \: \eta_1 \: + \: \frac{1}{(1-R){\bar{A_{II}}}}
\left[{({\bar{t}}-{\bar{t_{II}}})}^{1-R}-
{({\bar{t_1}-{\bar{t_{II}}})}}^{1-R}\right] \ \ \ \ \ \  
\bar{t_1}< \bar{t}<\bar{t_2}
\ee
Finally, the variable $\eta$ in the matter dominated stage is:
\be \label{et}
\eta \: = \: \eta_2 \: + \: \frac{1}{(1-M)A_{III}}
\left[t^{1-M} - \bar{t_2}^{1-M}\right] \ \ \ \ \ \ \ \ \  \bar{t_2}< t<t_0
\ee

By inverting the above relations eqs.(\ref{bbet}-\ref{et}) 
we obtain in the inflationary stage:
\bef
\bar{\bar{t}} \: = \: \bar{\bar{t_I}} - {\left[{(\bar{\bar{t_I}}-
\bar{\bar{t_i}})}^{Q+1} \: - \: (\bar{\bar{\eta}}-\eta_i)(Q+1)
\bar{\bar{A_{I}}}\right]}^{\frac{1}{Q+1}}
\eef
in the radiation stage
\bef
\bar{t} \: = \: \bar{t_{II}} \: + \: {\left[{(\bar{t_1}-\bar{t_{II}})}^{1-R} 
\: + \: (\bar{\eta}-\eta_1)(1-R){\bar{A_{II}}}\right]}^{\frac{1}{1-R}}
\eef
and in the matter dominated stage
\bef
t \: = \: {\left[{\bar{t_2}}^{1-M} \: + \: (\eta-\eta_2)(1-M){A_{III}}
\right]}^{\frac{1}{1-M}}
\eef

With suitable rearrangements of constants, we substitute the above
expressions on scale factor eqs.(\ref{Descr}) and we have the 
following minimal model written in conformal time:
\bes 
\bar{\bar{a_I}}(\bar{\bar{\eta}}) & = & {\alpha_I} {(\eta_I - {\bar{\bar
{\eta}}})}^{-q} \label{etagen1} \\
\bar{a_{II}}(\bar{\eta}) & = & {\alpha_{II}} {(\eta_{II} + \bar{\eta})}^r 
\nonumber \\
a_{III}(\eta) & = & {\alpha_{III}} {(\eta_{III} + \eta)}^m \nonumber
\ees
where the exponents of temporal dependences $q$, $r$, $m$
have the following expressions:
\bes \label{expo}
q & = & {Q\over{Q+1}} \ \ \ \  , \ \ \ \  r = {R \over{1-R}} \ \ \ \ , 
\ \ \ \  m = {M \over{1-M}} 
\ees
and the parameters $\alpha_j$, $\eta_j$ $(j=I,II,III)$ are: 
\bes \label{anul} 
\alpha_I & = & {\left(\bar{\bar{A_{I}}} {(Q+1)}^{-Q} \right)}^{1\over{Q+1}} 
\ \ \ \ , \ \ \ \ 
\eta_I =  \eta_i + \frac{{(\bar{\bar{t_I}}-\bar{\bar{t_i}})}^{Q+1}}
{(Q+1){\bar{\bar{A_{I}}}}}  \\
\alpha_{II} & = & {\left(\bar{A_{II}} {(1-R)}^{R}\right)}^{1\over{1-R}} 
\ \ \ \ , \ \ \ \
\eta_{II} =  \frac{{(\bar{t_1}-\bar{t_{II}})}^{1-R}}{(1-R){\bar{A_{II}}}}
\: - \: \eta_1 \nonumber \\
\alpha_{III} & = & {\left({A_{III}} {(1-M)}^M \right)}^{1\over{1-M}}
\ \ \ \ , \ \ \ \
\eta_{III} =  \frac{{\bar{(t_2)}}^{1-M}}{(1-M)A_{III}} \: - \: \eta_2 
\nonumber 
\ees
Expressions (\ref{anul}) are the most general relations among 
parameters of a scale factor description in conformal time
and the corresponding one in cosmic time. No assumptions and
no use of continuity conditions have been yet made.
When the scale factor in cosmic time description satisfies continuity
conditions, their properties eqs.(\ref{MAtch1})
allow us to write the relations for $\alpha_j$, $\eta_j$ as function 
of a more reduced number of parameters and
transition times $\bar{t_1}$ and $\bar{t_2}$:
\bes
\alpha_I & = & {\left(\bar{A_{II}}{\left({Q\over {R}}\right)}^Q {(Q+1)}^{-Q}
\ {\left(\bar{t_1}-(1-\frac{R}{M})\bar{t_2}\right)}^{R+Q} 
\right)}^{1\over{Q+1}} \label{set1} \\
\eta_I & = & \eta_i + \frac{{\left(\bar{t_1}(1+\frac{Q}{R})-
(\frac{Q}{R}-\frac{Q}{M})\bar{t_2} - \bar{\bar{t_i}} \right)}^{Q+1}}
{(Q+1){\bar{A_{II}}}{(\frac{Q}{R})}^Q{{\left(\bar{t_1}-(1-\frac{R}{M})
\bar{t_2}\right)}^{R+Q}}}  \nonumber  \\
\alpha_{II} & = & {\left(\bar{A_{II}} {(1-R)}^{R} \right)}^{1\over{1-R}}  
\ \ \ \ \ , \ \ \ \ \ 
\alpha_{III} =  {\left(\bar{A_{II}} {(1-M)}^M {\left({R\over M}\right)}^R
\ \ {\bar{t_2}}^{R-M}\right)}^{1\over{1-M}}  \nonumber \\
\eta_{II} & = & \frac{{\left(\bar{t_1}-(1-\frac{R}{M})\bar{t_2}\right)}^{1-R}}
{(1-R){\bar{A_{II}}}} \: - \: \eta_1 \ \ \ , \ \ \
\eta_{III}  =  \frac{{\bar{t_2}}^{1-R}}{(1-M){\bar{A_{II}}}{(\frac{R}{M})}^R}
\: - \: \eta_2 \nonumber 
\ees
meanwhile the conformal time variables (\ref{bbet}-\ref{et}) take the next 
expressions:
\bes
\bar{\bar{\eta}} & = & \eta_i + \left[\frac{{\left(\bar{t_1}(1+\frac{Q}{R})-
\bar{t_2}(\frac{Q}{R}-\frac{Q}{M})-\bar{\bar{t}}_i\right)}^{Q+1}}
{(Q+1){\left(\frac{Q}{R}\right)}^Q
\bar{A_{II}}{\left(\bar{t_1}-(1-\frac{R}{M})\bar{t_2}\right)}^{R+Q}}\right.
\: + \:  \label{BBEta} \\
& & \ \ \  - \: \left.\frac{{\left(\bar{t_1}(1+\frac{Q}{R})-
\bar{t_2}(\frac{Q}{R}-\frac{Q}{M})-\bar{\bar{t}}\right)}^{Q+1}}
{(Q+1){\left(\frac{Q}{R}\right)}^Q
\bar{A_{II}}{\left(\bar{t_1}-(1-\frac{R}{M})\bar{t_2}\right)}^{R+Q}}\right] 
\ \ \ \ \ \ \ \ \ \ \ \ \ \ \ \bar{\bar{t_i}}<\bar{\bar{t}}<\bar{t_1} 
\nonumber \\
\bar{\eta} & = & \eta_1 \; + \; \frac{\left[{\left(\bar{t}-(1-\frac{R}{M})
\bar{t_2}\right)}^{1-R} \; - \;
{\left(\bar{t_1}-(1-\frac{R}{M})\bar{t_2}\right)}^{1-R}\right]}
{(1-R)\bar{A_{II}}} \ \ \ \ \  \bar{t_1} < \bar{t} < \bar{t_2} \nonumber \\ 
\eta & = & \eta_2 \; + \; \frac{\left[t^{1-M} \; - \; \bar{t_2}^{1-M}\right]}
{(1-M)\bar{A_{II}}{(\frac{R}{M})}^R\;{(\bar{t_2})}^{R-M}} 
 \ \ \ \ \ \ \ \ \ \ \ \ \ \ \ \ \ \ \ \ \ \ \ \ \ \ \ \bar{t_2} < t < t_0 
\nonumber
\ees

At this point, we have not considered whatever or not the conformal time
variable itself is continuous. But as consequence
of the continuity of the scale factor written in cosmic 
time-type variables along
$\bar{t_1}$ and $\bar{t_2}$, we have yet:
\bes
\left. {\frac{d{\bar{\bar{\eta}}}}{d{\bar{\bar{t}}}}}\right|_{{\bar{t_1}}^-} 
& = & \left. {\frac{d{\bar{\eta}}}{d{\bar{t}}}}\right|_{{\bar{t_1}}^+} \ = \ 
\frac{1}{\bar{A_{II}}{\left(\bar{t_1}-(1-\frac{R}{M})\bar{t_2}\right)}^R} 
\label{con1}\\
\left. {\frac{d{\bar{\eta}}}{d{\bar{t}}}} \right|_{{\bar{t_2}}^-} & = & 
\left. {\frac{d{\eta}}{d{t}}} \right|_{{\bar{t_2}}^+} \ = 
\  \frac{{\left(\frac{R}{M}\bar{t_2}\right)}^{-R}}{\bar{A_{II}}} \nonumber
\ees

\subsection{The continuous $\eta$ case}

\margen Now, we can impose conditions relating the three conformal 
time variables eqs.(\ref{BBEta}) since we have the freedom in the 
parameters defined as $\eta_i$, $\eta_1$ and $\eta_2$. 

We will constrain the conformal time to be continuous along the transitions at 
$\bar{t_1}$ and $\bar{t_2}$, where the scale factor written in cosmic 
time-type variables have fully satisfactories continuity conditions. 
In this way, we will have the conformal time  
univocally defined at these transitions. 
\bes
\eta_1 & = & \bar{\eta}({\bar{t_1}}) \ \ \equiv \ \ \bar{\bar{\eta}}
({\bar{t_1}})  \label{Con1}\\
\eta_2 & = & \eta{({\bar{t_2}})} \ \ \equiv \ \ \bar{\eta}(\bar{t_2})
\nonumber
\ees

These conditions and eqs.(\ref{bbet}-\ref{et}) give two relations linking 
the parameters $\eta_i$ and the two transitions $\eta_1$ and $\eta_2$: 
\bes 
\eta_2 & = & \eta_1 \: + \: \frac{1}{(1-R){\bar{A_{II}}}}
\left[{({\bar{t_2}}-{\bar{t_{II}}})}^{1-R}-
{({\bar{t_1}-{\bar{t_{II}}})}}^{1-R}\right] \label{et2in} \\ 
\eta_1 & = & \eta_i \: + \: \frac{1}{(Q+1){\bar{\bar{A_{I}}}}} 
\left[{(\bar{\bar{t_I}}-\bar{\bar{t_i}})}^{Q+1}-
{(\bar{\bar{t_I}}-\bar{{t_1}})}^{Q+1}\right] \label{et1in}
\ees
Making use of the continuity properties of scale factor in cosmic time 
description (\ref{match1}) we can write eq.(\ref{et1in}) as:
\bes
\eta_1 & = & \eta_i \: + \: \frac{\frac{Q}{R}{(\bar{t_1}-\bar{t_{II}})}^{1-R}}
{(Q+1)\bar{A_{II}}}{\left[{\left(1+\frac{\bar{t_1}-\bar{\bar{t_i}}}
{\frac{Q}{R}(\bar{t_1}-\bar{t}_{II})}\right)}^{Q+1}\: -\: 1\right]} 
\label{Et1in}
\ees
and by using the matching eqs.(\ref{Match1}) and relations (\ref{Ti})
we can express them as functions of the observational 
transition times $t_r$ and $t_m$:
\bes
\eta_2 & = & \eta_1 \: + \: \frac{{(\frac{R}{M})}^{1-R}}{(1-R)\bar{A_{II}}}
({t_m}^{1-R}-{t_r}^{1-R}) \label{et1} \\
\eta_1 & = & \eta_i \: + \: \frac{\frac{Q}{R}
{\left(\frac{R}{M}t_r\right)}^{1-R}}{(Q+1)\bar{A_{II}}}
{\left[{\left(\frac{t_i-t_I}{t_r-t_I}\right)}^{Q+1}\: - \: 1 \right]} 
\nonumber
\ees

The remaining freedom is used by convenience to eliminate
the parameter $\eta_{II}$ in order to simplify the expression of the 
radiation dominated stage scale factor: $ \eta_{II} = 0 $.
From (\ref{anul}) is immediately obtained:
\be
\eta_1 = \frac{{(\bar{t_1}-\bar{t_{II}})}^{1-R}}{(1-R){\bar{A_{II}}}} 
\label{et1se}
\ee
thus, we have $\eta_1$ as a parameter univocally fixed in terms of parameters
of the minimal model scale factor. With eq.(\ref{et2in}) and (\ref{et1se}) we 
obtain the fixed expression for $\eta_2$:
\be
\eta_2 = \frac{{(\bar{t_2}-\bar{t_{II}})}^{1-R}}{(1-R){\bar{A_{II}}}} 
\label{et2se}
\ee
and the parameter $\eta_i$ is equally fixed in terms of the cosmic time-type
variables from eqs.(\ref{Et1in}) and (\ref{et1se}).
\be
\eta_i = \frac{{(\bar{t_1}-\bar{t_{II}})}^{1-R}}{(Q+1)\bar{A_{II}}}
{\left[\frac{R+Q}{R(1-R)} \: - \: \frac{Q}{R}
{\left(1+\frac{\bar{t_1}-\bar{\bar{t_i}}}
{\frac{Q}{R}(\bar{t_1}-\bar{t_{II}})}\right)}^{Q+1}\right]} \label{etise}
\ee 

Using again the matching expressions eqs.(\ref{Match1}) and the 
relations (\ref{Ti})
on eqs.(\ref{et1se}-\ref{etise}) we can
finally write the expressions relating the conformal time transitions
$\eta_1$, $\eta_2$ and $\eta_i$ with the observational standard values 
for the transitions at the beginning of radiation dominated stage $t_r$, 
beginning of matter dominated stage $t_m$ and beginning of inflationary 
stage $t_i$:
\bes 
\eta_1 & = & \frac{{\left(\frac{R}{M}\:t_r\right)}^{1-R}}{(1-R){\bar{A_{II}}}} 
\label{et1ob} \ \ \ \ \ , \ \ \ \ \ \ 
\eta_2  = \frac{{\left(\frac{R}{M}\:t_m\right)}^{1-R}}{(1-R){\bar{A_{II}}}} \\
\eta_i & = & \frac{{\left(\frac{R}{M}\:t_r\right)}^{1-R}}{(Q+1){\bar{A_{II}}}}
{\left[\frac{R+Q}{R(1-R)} \: - \: \frac{Q}{R}{\left(\frac{t_i-t_I}
{t_r-t_I}\right)}^{Q+1}\right]}  \nonumber
\ees

Eqs.(\ref{et1se}-\ref{etise}) or the corresponding ones
in terms of observational transition times eqs.(\ref{et1ob})
constitute the continuity and simplicity conditions for the 
conformal time variable chosen. From eqs.(\ref{BBEta})
with help of eqs.(\ref{Ti}),(\ref{Match1}) and (\ref{et1ob}), 
the continuous conformal time as function of transition
times $t_r$ and $t_m$ and proper cosmic time $t$ in each stage is:
\bes 
\bar{\bar{\eta}} & = & {{{({R\over M} t_r)}^{1-R}}\over{\bar{A_{II}}
(Q+1)}} \left[{{R+Q}\over{R(1-R)}}-{Q\over R} 
{\left({{t_I -t}\over{t_I-t_r}}\right)}^{Q+1}\right] 
\ \ \ \ \  t_i < t < t_r \label{etcont1} \\
\bar{\eta} & = & {{{({R\over M} t)}^{1-R}}\over{\bar{A_{II}}(1-R)}} 
\ \ \ \ \ \ \ \ \ \ \ \ \ \ \ \ \ \ \ \ \ \  \ \ \ \ \ \ \ \ \ 
\ \ \ \ \ \ \ \ \  t_r < t < t_m \nonumber \\
\eta & = & {{{({R\over M} t_m)}^{1-R}}\over{\bar{A_{II}}(1-M)}}
\left[{{R-M}\over{R(1-R)}} + {M\over R} {\left({t\over{t_m}}\right)}^{1-M}
\right] \ \ \ \ \ t_m < t < t_0 \nonumber 
\ees

With this continuous conformal time (\ref{etcont1}), the scale factor
takes the form:
\bes \label{etadef}
\bar{\bar{a_I}}(\bar{\bar{\eta}}) & = & {\alpha_I} {(\eta_I - {\bar{\bar
{\eta}}})}^{-q} \ \ \ \ \ \ \ \ \ \ \eta_i < \bar{\bar{\eta}} < \eta_1 \\
\bar{a_{II}}(\bar{\eta}) & = & {\alpha_{II}} {(\bar{\eta})}^r 
\ \ \ \ \ \ \ \ \ \ \ \ \ \ \ \ \ \eta_1 < \bar{\eta} < \eta_2  \nonumber  \\
a_{III}(\eta) & = & {\alpha_{III}} {(\eta_{III} + \eta)}^m \ \ \ \ \  
\eta_2 < \eta  \nonumber
 \ees
where the parameters $q$, $r$ and $m$ obey again relations (\ref{expo}) and
the coefficients $\alpha_j$ and  $\eta_j$ $(j=I,II,III)$
have these expressions since the continuity conditions (\ref{et1ob}) hold:
\bes \label{SET1}
\alpha_I & = & {\left(\bar{A_{II}}{\left({R\over {M}}\right)}^R 
{\left({Q\over {M}}\right)}^Q
{1 \over{{(Q+1)}^Q}} \; {t_r}^{R+Q} \right)}^{1\over{Q+1}} \\
\eta_I & = & {{R+Q}\over{R(Q+1)(1-R)}} {{{{({R\over M}t_r)}^{1-R}}}\over
{\bar{A_{II}}}} \nonumber \\
\alpha_{II} & = & {\bar{A_{II}}}^{1\over{1-R}} {(1-R)}^{R\over{1-R}} 
\ \ \ , \ \ \
\alpha_{III}  =  {\left(\bar{A_{II}} {(1-M)}^M {\left({R\over M}\right)}^R
{t_m}^{R-M}\right)}^{1\over{1-M}} \nonumber \\
\eta_{II} & = & 0 \ \ \ \ \  \ \ \ \ , \ \ \ \ \ \ \ \ \ \ \ \ \  \ \  
\eta_{III} =  {{M-R}\over{R(1-M)(1-R)}} 
{{{({R\over M} t_m)}^{1-R}}\over{\bar{A_{II}}}} \nonumber 
\ees

\subsection{Continuity of the scale factor in conformal time}

\margen The relations (\ref{etadef}) and (\ref{SET1}) have been 
obtained by  imposing to the conformal time
variable itself some conditions of simplicity and continuity at transtitions 
$\eta_1$ and $\eta_2$. The same results
could be reached by imposing equivalent relations to the scale factor
written in conformal time. 

We can take the more general case for the scale factor eqs.(\ref{etagen1}) and
impose it to have transitions at generic parameters $\eta_1$ and $\eta_2$ 
with continuity for the scale factor and its first derivative with respect to
conformal time. This matching led to the next four relations between the six
parameters of the scale factor $\alpha_j$, $\eta_{j}$ $(j = I, II, III)$:
\bes
\alpha_I & = & \alpha_{II} {\left(\frac{q}{r}\right)}^q {(\eta_1 + 
\eta_{II})}^{r+q} \label{MATch1}\\
\eta_I & = & \eta_1 \left(1+\frac{q}{r}\right) + \frac{q}{r} \; \eta_{II} 
\label{MATch2} \\
\alpha_{III} & = & \alpha_{II} {\left(\frac{m}{r}\right)}^{-m} {(\eta_{II}+
\eta_2)}^{r-m} \label{MATch3} \\
\eta_{III} & = & \eta_2 {\left(\frac{m}{r}-1\right)} + \frac{m}{r} \; 
\eta_{II} \label{MATch4}
\ees

Now, we find the conditions to be satisfied by the parameters $\eta_1$ and 
$\eta_2$ in order
to satisfy the matching conditions eqs.(\ref{MATch1}-\ref{MATch4}) when the 
conformal time
variable has been obtained from a cosmic time-type description with continuous
 and smooth transitions. Thus, we have in principle the general 
expressions (\ref{set1}) for $\alpha_j$, $\eta_j$ $(j=I,II,III)$.
It is straightforward to see that
eq.(\ref{MATch1}) is trivially satisfied. From eq.(\ref{MATch2}) we obtain:
\bes \label{sor1}
\eta_i & = & \eta_1  +  \frac{\frac{Q}{R}{(\bar{t_1}-
(1-\frac{R}{M})\bar{t_2})}^{1-R}}
{(Q+1)\bar{A_{II}}}{\left[1 -  {\left(1+\frac{\bar{t_1}-\bar{\bar{t_i}}}
{\frac{Q}{R}(\bar{t_1}-(1-\frac{R}{M})\bar{t_2})}\right)}^{Q+1}\right]} \ \ \ 
\ees
Meanwhile, both eqs.(\ref{MATch3}) and (\ref{MATch4}) give the same
condition:
\bes \label{sor2}
\eta_2 & = & \eta_1 \: + \: \frac{1}{(1-R){\bar{A_{II}}}}
\left[{\left(\frac{R}{M}{\bar{t_2}}\right)}^{1-R} \; - \;
{\left(\bar{t_1}-(1-\frac{R}{M})\bar{t_2}\right)}^{1-R}\right]
\ees
It is easy to see that eqs.(\ref{sor1}) and (\ref{sor2}) are equivalent to 
expressions (\ref{Et1in})
and (\ref{et2in}). By making use of expressions (\ref{Match1}) and 
(\ref{Ti}) we can write them as function of observational values:
\besf
\eta_2 \: - \: \eta_1 & = &  \frac{{(\frac{R}{M})}^{1-R}}{(1-R)\bar{A_{II}}}
({t_m}^{1-R}-{t_r}^{1-R})  \\
\eta_i \:  - \: \eta_1 & = & \frac{\frac{Q}{R}
{\left(\frac{R}{M}t_r\right)}^{1-R}}{(Q+1)\bar{A_{II}}}
{\left[1 \: - \: {\left(\frac{t_i-t_I}{t_r-t_I}\right)}^{Q+1} \right]} 
\eesf
That means, we have recovered the conditions (\ref{et1}) 
obtained from imposing 
continuity to conformal time variable itself before simplifying  $\eta_{II}=0$.

In fact, the four conditions of continuity of the scale factor written in
conformal time (\ref{MATch1}) are not independent 
equations when
the conformal time variable is constructed over a continuous description in 
cosmic time. Continuity in the first derivative of the conformal time
scale factor is warranted 
from continuity in the first derivative of the cosmic time description 
because equality of the
relative derivatives eqs.(\ref{con1}). Thus, the 
only two independent
conditions are the equality in the scale factor at the transitions $\eta_1$ 
and $\eta_2$. Satisfy 
them impose not only the continuity in the scale factor written in cosmic 
time at $\bar{t_1}$
and $\bar{t_2}$ but also the continuity in the conformal time variable itself 
along the
transitions, exactly as made in eqs.(\ref{Con1}) and 
remaining always one
freedom that we fix by imposing the simplifying condition $\eta_{II} = 0$.

\subsection{Summary of the scale factor in conformal time}

\margen We have used the scale factor description written in cosmic time-type 
variables (See eqs.(\ref{Descr}) and related ones) in order to define a
continuous conformal time variable and also the corresponding
scale factor with suitable continuity and simplicity properties:
\bes \label{EDescr} 
\bar{\bar{a_I}}(\bar{\bar{\eta}}) & = & {\alpha_I} {(\eta_I - {\bar{\bar
{\eta}}})}^{-q} \ \ \ \ \ \ \ \ \ \ \eta_i < \bar{\bar{\eta}} < \eta_1 \\
\bar{a_{II}}(\bar{\eta}) & = & {\alpha_{II}} {(\bar{\eta})}^r 
\ \ \ \ \ \ \ \ \ \ \ \ \ \ \ \ \ \eta_1 < \bar{\eta} < \eta_2 \nonumber \\
a_{III}(\eta) & = & {\alpha_{III}} {(\eta_{III} + \eta)}^m \ \ \ \ \  
\eta_2 < \eta  \nonumber
\ees
with continuous, sudden and smooth transitions (continuity of scale factor and
first derivative with respect to conformal time) at $\eta_1$ and $\eta_2$.
The parameters $\alpha_j$, $\eta_j$ $(j= I,II,III)$ satisfy these
matching relations:
\bes
\alpha_I & = & \alpha_{II} {\left(\frac{q}{r}\right)}^q {(\eta_1 
)}^{r+q} \label{MAT1} \ \ \ \ , \ \ \ \
\eta_I  =  \eta_1 \left(1+\frac{q}{r}\right)  \\
\alpha_{III} & = & \alpha_{II} {\left(\frac{m}{r}\right)}^{-m} 
{(\eta_2)}^{r-m} \ \ \ \ , \ \ \ \ 
\eta_{III} =  \eta_2 {\left(\frac{m}{r}-1\right)} \nonumber
\ees

It must be remembered that all the parameters of this scale factor in
conformal time have complete and univocally translation as functions
of the observational transition times $t_r$, $t_m$, the cosmic time
$t = c {\mathcal{T}}$ and the global scale factor constant $\bar{A_{II}}$.
This dictionary is constituted by the expressions given for the exponents 
$q$, $r$ and $m$ in eqs.(\ref{expo}) and the corresponding ones for 
$\alpha_j$, $\eta_j$ $(j=I,II,III)$ by eqs.(\ref{SET1}). 
Remember also the expressions for conformal time variables eqs.(\ref{etcont1})
and the expressions for the transition parameters $\eta_1$, $\eta_2$ and
$\eta_i$ eqs.(\ref{et1ob}). 

In this way, we have obtained a minimal model with continuous, sudden
and smooth transitions in cosmic time description (\ref{Descr}) and also in 
conformal time variables (\ref{EDescr}).

\section{Another Conformal Time Construction}
\setcounter{equation}0

\margen We have used above the descriptive cosmic time
variables in order to linking the conformal time variable with
the observational information.  Another alternative arises: it
is possible to link the conformal time variable {\it directly} with
the current cosmic time variable $(t)$ without passing the 
descriptive variables. 

   This possibility will give us a conformal time description
{\it closer} to cosmic time and observational information. So close
that it will hereditate the lacking of full satisfactory continuity
conditions, as we will see. Differently from the former one,
this conformal time variable satisfies continuity itself in
the proper time derivatives.
Transitions are studied directly on the observational values
$t_r$ and $t_m$ instead of the descriptive ones $\bar{t_1}$,
$\bar{t_2}$ where fully continuity could be introduced.
Scale factor in this conformal time variable is not as continuous
as last case. Their transitions can be made sudden and
continuous, but never smooth. Matching relations for this
scale factor are differents from the above given.
For studies where full continuity of scale factor can be overcome or
closeness to observational information is prioritary, this approach
could result useful.
\pagebreak[4]

Again, we will have a particular conformal time variable for each stage, 
oughted to the different shape of the scale factor . We define:
\bes
d \bar{\bar{\eta}} & = & \frac{ d{t'}}{\bar{\bar{a_I}}(t')} \ \
\ \ \ \ \ {t_i}  <  t' < {t_r}  \label{Etain}  \\
d \bar{\eta} &  =  & \frac{d{t'}}{\bar{a_{II}}(t')}  \ \ \ \ \  
{t_r} < {t'}  < {t_m}  \nonumber \\
d \eta & = & \frac{d{t'}}{a_{III}(t')} \ \ \ \
{t_m} <  t' < {t_0}   \nonumber
\ees
where $a_I(t)$, $a_{II}(t)$ and $a_{III}(t)$ are given by 
eqs.(\ref{Dest}).
Working as made in the general conformal time case, we integrate the equations
above and obtain expressions
for the conformal time variable in each stage. Thus, we define the variable
${\bar{\bar{\eta}}}$ for the inflationary stage, $\bar{\eta}$ for radiation
dominated stage and $\eta$ for matter dominated stage:
\besf
{\bar{\bar{\eta}}} & = & \eta_i + \frac{{\left(\frac{t_I}{t_r}-1\right)}^{-Q}}
{(Q+1){\bar{A_{II}}}{{t_r}^{R+Q}}
{(\frac{R}{M})}^R} \left[{(t_I-t_i)}^{Q+1}-
{(t_I-t)}^{Q+1}\right] \ \ \ \ \ \  t_i<t<t_r \\
\bar{\eta} & = &\eta_1 + \frac{1}{(1-R){\bar{A_{II}}}{(\frac{R}{M})}^R}
(t^{1-R} - {t_r}^{1-R}) \ \ \ \ \ \ \ \ \ \ \ \ \ \ \ \ \ \ \ \ \ \ \ \  
t_r<t<t_m \\
\eta & = &\eta_2 + \frac{1}{(1-M){\bar{A_{II}}}{{t_m}^{R-M}}
{(\frac{R}{M})}^R}(t^{1-M} - {t_m}^{1-M}) \ \ \ \ \ \ \ \ \ \ t_m<t<t_0
\eesf
where we have defined this time the parameters
$\eta_i =\bar{\bar{\eta}}(t_i)$, 
$\eta_1 = \bar{\eta}(t_r) $ and  $\eta_2 = \eta(t_m)$.
By inverting the above relations for $\eta$, ${\bar{\eta}}$ and
$\bar{\bar{\eta}}$ we obtain expressions for $t$ in each one of
the stages:
\besf
t_I - t  & = & {\left[{(t_I-t_i)}^{Q+1} - (\bar{\bar{\eta}}-\eta_i)(Q+1)
\bar{A_{II}} {\left(\frac{R}{M}\right)}^R {t_r}^{R+Q} 
{\left(\frac{t_I}{t_r}-1\right)}^Q 
\right]}^{\frac{1}{Q+1}} \  t_i<t <t_r \\
t & = &{\left[{t_r}^{1-R} + (\bar{\eta}-\eta_1)(1-R){\bar{A_{II}}}
{\left(\frac{R}{M}\right)}^R\right]}^{\frac{1}{1-R}}
\ \ \ \ \ \ t_r < t < t_m \\
t & = &{\left[{t_m}^{1-M} + (\eta-\eta_2)(1-M){\bar{A_{II}}}
{\left(\frac{R}{M}\right)}^R {t_m}^{R-M} \right]}^{\frac{1}{1-M}}
t_m<t<t_0
\eesf

With some straightforward computations 
and suitable rearrangements of constants, we obtain expressions for
the scale factor as seen in eqs.(\ref{etagen1}) whose temporal 
dependence exponentes are
given again by eqs.(\ref{expo}) and the parameters
$\alpha_j$, $\eta_j$ $(j=I,II,III)$ are: \pagebreak[4]
\bes
\alpha_I & = & {\left(\bar{A_{II}}{\left({R\over {M}}\right)}^R 
{\left({\frac{t_I}{t_r}-1} \right)}^Q {(Q+1)}^{-Q} \ \ {t_r}^{R+Q} 
\right)}^{1\over{Q+1}} \\
\eta_I & = & \eta_i + \frac{1}{(Q+1){\bar{A_{II}}}{{t_r}^{R+Q}}
{(\frac{R}{M})}^R}{\left(\frac{t_I}{t_r}-1\right)}^{-Q} {(t_I-t_i)}^{Q+1} 
\nonumber \\
\alpha_{II} & = & {\left(\bar{A_{II}} {(1-R)}^{R}
{\left(\frac{R}{M}\right)}^{R}\right)}^{1\over{1-R}} \ \  , \ \  
\alpha_{III}  =  {\left(\bar{A_{II}} {(1-M)}^M {\left({R\over M}\right)}^R
\ \ {t_m}^{R-M}\right)}^{1\over{1-M}} \nonumber \\
\eta_{II} & = & \frac{{t_r}^{1-R}}{(1-R){\bar{A_{II}}}{(\frac{R}{M})}^R}
- \eta_1 \ \ \ \ \ \ , \ \ \ \ \ \
\eta_{III}  =  \frac{{t_m}^{1-R}}{(1-M){\bar{A_{II}}}{(\frac{R}{M})}^R}
- \eta_2 \nonumber
\ees

Now, we can impose conditions relating the three conformal time variables 
in order to eliminate the freedom in the parameters defined as $\eta_i$,
$\eta_1$ and $\eta_2$. We use it for constraining the conformal time to
be continuous along the transitions at $t_m$ and $t_r$ (and, as consequence,
univocally defined at these transitions).
\besf
\eta_1 & = & \bar{\eta}({t_r}) \ \ \equiv \ \ \bar{\bar{\eta}}({t_r}) \\
\eta_2 & = & \eta{({t_m})} \ \ \equiv \ \ \bar{\eta}({t_m})
\eesf
and the remaining freedom is used again in eliminate
the parameter $\eta_{II}$, simplifying thus the expression of the 
radiation dominated stage scale factor $\eta_{II} = 0$.
Whit these three conditions,
we can fix the parameters $\eta_i$, $\eta_1$ and $\eta_2$ as functions
of parameters of the minimal model only. The parameter $\eta_i$ is:
\be
\eta_i = \frac{1}{{t_r}^R {\bar{A_{II}}} {(\frac{R}{M})}^R}
\left[{\frac{t_r}{(1-R)}}-{\frac{(t_I-t_r)}{(Q+1)}}\left[{\left({\frac
{t_I-t_i}{t_I-t_r}}\right)}^{Q+1}-1\right]\right]
\ee
and the parameters related to the transitions are: 
\bes
\eta_1 & = & {{{t_r}^{1-R}}\over{\bar{A_{II}} (1-R) {({R\over M})}^R}}
 \ \ \ \ , \ \ \ \ \ 
\eta_2  =  {{{t_m}^{1-R}}\over{\bar{A_{II}} (1-R) {({R\over M})}^R}}  
\ees
In this way, the conformal time variables are continuous and univocally 
defined at transitions
$\eta_1$ and $\eta_2$. In terms of minimal model parameters, they are:
\bes \label{Eperf}
\bar{\bar{\eta}} & = & {{ {(t_r)}^{-R}}\over{\bar{A_{II}}
{\left({R\over {M}}\right)}^R}(Q+1)}
\left[{{R+Q}\over{(1-R)}}\ t_r + {t_I} -
{{{(t_I -t)}^{Q+1}\over{(t_I-t_r)}^Q }}\right] \\
\bar{\eta} & = & {{{(t)}^{1-R}}\over{\bar{A_{II}}(1-R)
{\left({R\over M}\right)}^R}} \nonumber \\
\eta & = & {{ {(t_m)}^{1-R}}\over{\bar{A_{II}}(1-M)
{\left({R\over M}\right)}^R}} 
\left[{{R-M}\over{1-R}} + {\left({t\over{t_m}}\right)}^{1-M}
\right] \nonumber
\ees
Notice this conformal time variable enjoys continuity along the
transitions and observational times $t_r$ and $t_m$, not only
on the variable itself but also in their derivative with respect
to proper cosmic time in each stage:
\bes
\left. {\frac{d{\bar{\bar{\eta}}}}{dt}}\right|_{{t_r}^-} & = & 
\left. {\frac{d{\bar{\eta}}}{dt}}\right|_{{t_r}^+} \ = \ 
\frac{{t_r}^{-R}}{\bar{A_{II}}{\left(\frac{R}{M}\right)}^R} 
\label{Econ1}\\
\left. {\frac{d{\bar{\eta}}}{dt}} \right|_{{t_m}^-} & = & 
\left. {\frac{d{\eta}}{d{t}}} \right|_{{t_m}^+} \ = \ 
\frac{{t_m}^{-R}}{\bar{A_{II}}{\left(\frac{R}{M}\right)}^R} 
\nonumber
\ees
 
With these variables, we obtain the model written in continuous 
conformal time
as can be seen in eqs.(\ref{etadef}). In this case, its parameters take this
form:
\bes \label{Eset1}
\alpha_I & = & {\left(\bar{A_{II}}{\left({R\over {M}}\right)}^R 
{\left({\frac{t_I}{t_r}-1} \right)}^Q {(Q+1)}^{-Q} \ \ {t_r}^{R+Q} 
\right)}^{1\over{Q+1}} \\
\eta_I & = & \frac{{t_r}^{1-R}}{{\bar{A_{II}}}(Q+1){\left(
{R\over M}\right)}^R} \left[\frac{t_I}{t_r}+\frac{Q+R}{1-R}\right] 
\nonumber \\ 
\alpha_{II} & = & {\left(\bar{A_{II}} {(1-R)}^{R} 
{\left({R\over M}\right)}^R\right)}^{\frac{1}{1-R}} \ \  ,  \ \
\alpha_{III}  =  {\left(\bar{A_{II}} {(1-M)}^M {\left({R\over M}\right)}^R
{t_m}^{R-M}\right)}^{1\over{1-M}} \nonumber \\
\eta_{II}  & = & 0 \ \ \ \ \  , \  \ \ \ \ \ \ \ \ \ \ \
\eta_{III} = {{M-R}\over{(1-M)(1-R)}}\:  
{{{(t_m)}^{1-R}}\over{\bar{A_{II}}{\left({R\over {M}}\right)}^R}} \nonumber
\ees
Notice that parameters (\ref{Eset1}) do not satisfy eqs.(\ref{MAT1}).
That means, the scale factor  (\ref{EDescr}) does not have
sudden, continuos and smooth transitions among stages when
it runs on this conformal time variables (\ref{Eperf}). Scale factor
is continuous, but not thus its first derivative with respect to conformal
time. In this case, we have sudden and continuous transitions at
$\eta_1$ and $\eta_2$, but they are not smooth. 

In fact, we have continuity in the scale factor 
$\bar{\bar{a_I}}(\eta_1) = \bar{a_{II}}(\eta_1)$ and
$\bar{a_{II}}(\eta_2) = a_{III}(\eta_2)$. For the first
derivative, instead of  equality among both sides of
transition, we have:
\besf
\left. {\frac{d{\bar{\bar{a_I}}}(\bar{\bar{\eta}})}
{d{\bar{\bar{\eta}}}}}\right|_{{\eta_1}^-} & = & 
\frac{R}{Q}\left(\frac{t_I}{t_r}-1\right)
\left. {\frac{d{\bar{a_{II}}}(\bar{\eta})}
{d{\bar{\eta}}}}\right|_{{{\eta_1}}^+} \\
\left. {\frac{d{\bar{a_{II}}}(\bar{\eta})}
{d{\bar{\eta}}}} \right|_{{{\eta_2}}^-} & = & 
\frac{M}{R}
\left. {\frac{d{a_{III}}(\eta)}{d{\eta}}} \right|_{{{\eta_2}}^+}
\eesf
Explanation is simple: this conformal time variable maps
very closer the cosmic proper times. So closer that
equality in its derivatives with respect to proper cosmic times 
hold along the transitions. But we know there is an inherent
discontinuity in the proper cosmic time derivatives of
scale factor, as discussed above (See eqs.(\ref{Dest}-\ref{conti}) and 
comments below). This discontinuity
was absorbed in the descriptive cosmic time variables, as can
be seen from their derivatives eqs.(\ref{tat}).
Lacking of this descriptive variables, this construction of
conformal time is not able to absorb such discontinuity at
transitions $t_r$ and $t_m$ and the related proportionality
factors (\ref{tat})  reappear in the conformal time derivatives of
scale factor.

By arranging in the right way the equality among the right-hand
and left-hand derivatives, the matching equations for the scale
factor (\ref{EDescr}) with conformal time (\ref{Eperf}) 
take the form:
\bes
\alpha_I & = & \alpha_{II}{(\eta_1)}^{r+q}{\left(\frac{q}{r}\right)}^q
{\left(\frac{Q}{R}\frac{t_r}{t_I-t_r }\right)}^{-q} \ \ , \ \ 
\eta_I  =  \eta_1 \left(1+\frac{q}{r}{\left(\frac{Q}{R}
\frac{t_r}{t_I-t_r}\right)}^{-1}\right) \\
\alpha_{III} & = & \alpha_{II}{\eta_2}^{r-m} {\left(\frac{m}{r}
\frac{R}{M}\right)}^{-m}  \ \ \ \ \ \ \ \ , \ \ \ \ \ \ \ \ 
\eta_{III}  =  \eta_2 {\left(\frac{m}{r}\frac{R}{M}-1 \right)} \nonumber
\ees

\section{Conclusions}

\margen Our starting point and motivation was to realize that modelized 
transitions can not reproduce real transitions. 
We point out that in cosmologies dealing with discountinuous
stages describing a step-by-step evolution for the scale factor
without underlying satisfactory description of transitions,
is not enough to match the scale factor. At priori, the
spacetime metric can be discontinuous, leading to
discontinuities in the temporal variables itself.

As the description of different stages come from different coordinate systems,
there exists the possibility of transformations relating them.
We have used this freedom choosing coordinates with satisfactory
continuous matching along transitions and containing the minimal
information about the observational Universe.
 
The minimal observational information are the transition time scales. It
gives us scale factor ratii reached in each evolution stage as a powerful tool.
All our constructions are constrained to reach the same expansion ratii
in each one of the stages eqs.(\ref{rin}-\ref{rmat}). 

The set of parameters relating descriptive scale factor and observational 
Universe information is given. This scale factor (\ref{Descr}) runs on  
transformed variables ((\ref{tbar}),(\ref{tbarbar}))
and has modified durations on inflationary and radiation dominated stages.
The matter dominated stage is considered untransformed in the
description, since observations must be made from this current stage.
Modelized transitions happen at transformed temporal coordinates,
related with the observational values (\ref{Ti}).

Also, a set of parameters giving satisfactory continuous and predictive
conformal time minimal model is computed. The conformal time
variable (\ref{etcont1}) is constructed on the descriptive model and 
constrained 
to be continuous along the transformed transitions (\ref{et1ob}). 
A remaining 
freedom is invested in simplicity for the corresponding scale factor 
expressions (\ref{EDescr}). 

The minimal model thus obtained have continuous, sudden and
smooth transitions both in cosmic time and in conformal time
descriptions. Minimal information of observational Universe have
been introduced in both cases. Parameters and variables are linked with 
observational standard values and current proper cosmic time in a totally 
determined way
((\ref{Match1}),(\ref{expo}),(\ref{MAT1}) and related expressions).
It preserves the physical 
meaning and predictability of computations made with these constructions
and allows them to be confronted with measurementes and observations.
For instance, gravitational wave generation and related type computations
can benefit of this kind of treatment.

We have included also another conformal
time variable with different continuity and predictive characteristics.
It is constructed on the proper cosmic time directly and maps it 
closer than the former one (\ref{Eperf}). This variable is smoothly continuous
along the transitions at observational times.  In opposite way, the
scale factor running on it presents sudden and continuous transitions,
but never smooth. It is oughted to the lack of descriptive intermediate
variables absorbing the inherent discontinuity in the proper time
derivatives of scale factor at real transitions.

Of course, we do not consider this treatment solves the problem of
transitions among the stages for which a full dynamical description is
required. We recall the problem of continuity
of temporal coordinate variable in step-by-step cosmologies 
and we suggest some ways in order to overcome it, at least
useful in extraction of some observational consequences.
It is clearly useful in metric perturbations type computations,
and it can be applied in whatever subject involving different 
cosmological evolution stges.
It opens the way to new and more exact treatments.

\section*{Appendixes}

\appendix
\section{Scale Factor Description for Power Law Inflationary Stages}
\setcounter{equation}0

\margen In the last sections we have treated a minimal model including an
inverse power inflationary stage. We recall that this process is
easily genralized to other type of inflationary stages. In sake of
completeness, we report here the results and link with observational
information for a model including an arbitrary power law
inflationary stage. Notice some differences in the inflationary scale factor
would modify expressions in the linking process.

The minimal model to be linked with observational information
would be in this case:
\bes \label{pdescr}
\bar{\bar{a_I}}(\bar{\bar{t}}) & = & \bar{\bar{A_{I}}}
{(\bar{\bar{t}}-\bar{\bar{t_I}})}^{-Q} \ \  \ \ \ \ \ \ \ 
{\bar{\bar{t}}} < \bar{t_1}\\
\bar{a_{II}}(\bar{t}) & = & \bar{A_{II}} {(\bar{t}-\bar
{t_{II}})}^R  \nonumber \ \ \ \ \ \ \  {\bar{t_1}} < {\bar{t}} < {\bar{t_2}} \\
a_{III}(t) & = & A_{III} {(t - t_{III})}^M  \ \ \ \ \ \ \ \  
{\bar{t_2}} < t\nonumber
\ees
The inflationary stage $(I)$ is an accelerated expansion provided $Q>1$
and $t_I< t_i<t_r$. The usual power law is found for $t_I=0$. The same
behaviours before considered for radiation and matter dominated stages
hold. Again, we impose this model to have continuous transitions
at $\bar{t_1}$ and $\bar{t_2}$ both for scale factor and first derivative.
The matching relations thus obtained are:
\bes 
\bar{\bar{t_I}} & = & \bar{t_1} \left(1-{Q \over{R}}\right) + \bar{t_{II}} 
{Q \over{R}} \label{pmatch1} \ \ \  , \ \ \ 
t_{III}  =  \left(1 - {M \over{R}}\right) \bar{t_2} + {M \over{R}} 
\bar{t_{II}}  \\
\bar{\bar{A_I}} & = & \bar{A_{II}} {\left({R \over{Q}}\right)}^Q {(\bar{t_1} - 
\bar{t_{II}})}^{R-Q} \ \ \  , \ \ \ 
A_{III} =  \bar{A_{II}} {\left({R \over{M}}\right)}^{M} {(\bar{t_2} - 
\bar{t_{II}})}^{R-M} \nonumber 
\ees
In the linking with observational Universe information, we apply again
the relations (\ref{rin}-\ref{rmat}) and fix also temporal coordinate
in the third stage with the corresponding proper cosmic time as made
in (\ref{tiii}). With this, we obtain the expressions for the transitions:
\bes  
\bar{t_1} & = & {R \over{M}} t_r + \left(1 - {R \over{M}}\right) t_m 
\ \ \ \ \ \ , \ \ \ \ 
\bar{t_2}  =  t_m \label{pTi} \\
\bar{\bar{t_i}} & = &
 \left({R \over{M}} - {Q\over{M}}{{t_r-t_i}\over{t_r-t_I}} \right)t_r
+ \left(1 - {R \over{M}}\right) t_m \nonumber
\ees
and the descriptive variables are:
\bes \label{ptbar}
\bar{\bar{t}} &  =  &
 \left({R \over{M}} - {Q\over{M}}{{t_r-t}\over{t_r-t_I}} \right)t_r
+ \left(1 - {R \over{M}}\right) t_m  \\
\bar{t} & = & {R \over{M}} t + \left(1 - {R \over{M}}\right) t_m 
\nonumber
\ees
In terms of observational transition values $t_r$ and $t_m$, the
parameters of the scale factor description take the form:
\bes \label{pMatch1}
\bar{\bar{t_I}} & = & {t_r} {\left({R\over{M}}-{Q\over{M}}\right)} + {t_m}
{\left({1 - {R \over{M}}}\right)} \ \ \ , \ \ \  
\bar{t_{II}}  =  \left({1-{R\over{M}}}\right) t_m  \\
\bar{\bar{A_I}} & = & \bar{A_{II}} {\left({M \over{Q}}\right)}^Q 
{\left({R \over{M}}\right)}^R {t_r}^{R-Q} \ \ \ \ \ \ \ , \ \ \  \
A_{III}  =  \bar{A_{II}} {\left({R \over{M}}\right)}^{R} {t_m}^{R-M} 
\nonumber \ees

In this case, the duration of stages is again modified in the descriptive
variables. We have matter dominated hold the same duration. For
radiation dominated and inflationary stages we have:
\besf
\left.{\Delta \bar{t}}\right|_{RAD} & = & \frac{R}{M} 
\left.{\Delta t}\right|_{RAD}  \\
\left.{\Delta {\bar{\bar{t}}}}\right|_{INF} & = & \frac{Q}{M}
\left(1+\frac{t_I}{t_r-t_I}\right) \left.{\Delta t}\right|_{INF}
\eesf
In this case, the radiation dominated stage is again contracted
in the descriptive variables. For inflationary stage, a dilatation
takes place in whatever case. For pure power law, we have
${\Delta {\bar{\bar{t}}}} =  \frac{Q}{M}{\Delta t}$ where
$\frac{Q}{M}>1$ for inflationary behaviours.

\section{An Example of Intermediate Transition Stage}
\setcounter{equation}0

\margen As an example, we present a more realistic alternative to the 
transitions here supposed sudden. We attempt to elaborate
a continuous and smooth evolution of scale factor
running on a proper cosmic time coordinate fixed
before, along and after the transition. 

Let us consider an intermediate stage
between the radiation dominated and matter dominated ones. 
As above said, it is not possible a sudden and smoothly continuous 
transition there. We relaxe these assumptions by considering the intermediate
brief stage $a_{tr}(t)$ with variable evolution law $N(t)$ matching 
continuous and smoothly with the radiation stage at $t_m$
and with the matter stage at $t_m + \Delta$:
\bef
a_{II}(t) = A_{II} t^R \ \ \ \ \ \ \ a_{tr}(t) = A_{II} t^{N(t)} 
\ \ \ \ \ \ a_{III}(t) = t^M
\eef
In the exit of radiation dominated stage $t_m$, the conditions to be 
satisfied are:
\be \label{exr}
a_{II}(t_m) = a_{tr}(t_m)  \ \ \   , \ \ \
\left. \dot{a_{II}}(t)\right|_{t_m} = \left. \dot{a_{tr}}(t)\right|_{t_m}
\ee
where dot means derivative with respecto to the fixed proper
cosmic time coordinate $t$.
In the beginning of matter dominated stage at $t_m+\Delta$ we have:
\be \label{bma}
a_{tr}(t_m+\Delta) =  a_{III}(t_m+\Delta) \ \  \ , \ \ \
\left. \dot{a_{tr}}(t)\right|_{t_m+\Delta} = 
\left. \dot{a_{III}}(t)\right|_{t_m+\Delta}
\ee
From eqs.(\ref{exr}), we obtain two boundary conditions for the
function $N(t)$ in the intermediate stage: $N(t_m) = R$ and
$\left. \dot{N(t)}\right|_{t_m} = 0$. From eqs.(\ref{bma}), we
extract a relation among the constants of global scale factors:
\bef
A_{III} = A_{II} {\left(t_m+\Delta \right)}^{N(t_m+\Delta)-M}
\eef
Notice this is a different relation from obtained in the
treatment with descriptive variables eq.(\ref{Match1}).
Finally, the remaining information is:
\be \label{deri}
\left. \dot{N(t)}\right|_{t_m+\Delta} \ln\left(t_m+\Delta \right) =
\left(M-N\left(t_m+\Delta \right)\right) {\left(t_m+\Delta \right)}^{-1}
\ee
Without further information about $\Delta$, this equation gives
us as possible solution: $N\left(t_m+\Delta \right)=M$, 
$\left. \dot{N(t)}\right|_{t_m+\Delta} = 0$ and $ A_{III} = A_{II}$.
Thus, $N(t)$ would be a function interpolating
among values $R$ and $M$ in a brief interval $\Delta$ with
vanishing derivative in both extremes. This behaviour would mean a not
negligible dynamics of scale factor during the transition, with a changing
evolution in a short interval.

If we consider an arbitrary function $N(t)$ without vanishing derivatives
at the extremes, the eq.(\ref{deri}) can be
approximated in the case of smooth interpolation leading to the
next relation:
\be
N\left(t_m+\Delta \right) = R \frac{A}{A+1} + M \frac{1}{1+A}
\ee
where $A= \left(1+\frac{t_m}{\Delta}\right) \ln(t_m+\Delta)$.
In the approximation of transition almost sudden, we can take $\Delta 
\rightarrow \delta$ neglecting quadratic terms. In this case, we
obtain $N(t_m+\delta) \sim R + (M-R)\frac{\delta}{t_m}\ln(t_m)$
and as consequence $A_{III}\sim A_{II} 
{(t_m+\delta)}^{(R-M)(1-\frac{\delta}{t_m}\ln t_m)}$. We recover
the expressions like (\ref{MAtch1}).  Notice in anycase that
more complex dynamics are in principle possible and without
better comprehension of phenomena driving the transitions, it is
not possible to give an unique and satisfactory description of them.

\section{Scale Factor Description Based on Radiation Dominated Stage}
\setcounter{equation}0

\margen Until now, we have worked with minimal models for scale factor
evolution including the current matter dominated stage. Link with
observations made from this last stage have advised us to 
privilege it in conserving its proper cosmic time coordinate
in our description. Again, alternatives arise. It is possible privilege
the proper cosmic time of radiation dominated stage, if one
wants consider only the first stages in minimal model or not
scaling of radiation dominated stage is preferred. Obviously,
the needed rescaling of variables and durations acquite in this
case the current stage and current time, from where observational
information is provided.
 
\pagebreak[4]
The scale factor in descriptive variables for this choice is:
\bes \label{RDescr}
\bar{\bar{a_I}}(\bar{\bar{t}}) & = & \bar{\bar{A_{I}}}
{(\bar{\bar{t_I}}-\bar{\bar{t}})}^{-Q} \ \ \ \ \ \ \ 
{\bar{\bar{t_i}}}  <  {\bar{\bar{t}}} < {t_r}   \\
\bar{a_{II}}(\bar{t}) & = & \bar{A_{II}} {(\bar{t})}^R 
 \ \ \ \ \ \ \ \ \ \ \    {t_r} < \bar{t}  < {t_m}\nonumber \\
a_{III}(t) & = & A_{III} {(\bar{t}-\bar{t_{III}})}^M 
\ \ \ \ \ \ \  \  {t_m} <  t < \bar{t_0} \nonumber
\ees
where  the radiation dominated stage runs on its proper
cosmic times among the proper transition times (that we
have called also $t_r$ and $t_m$). Descriptive variables 
are introduced in inflationary stage and in the current
matter dominated stage. The scale factor has
continuous and smooth transitions
at $t_r$ and $t_m$ and each stage satisfy the proper
expansion ratii (\ref{rin}-\ref{rmat}).
The parameters
of scale factor (\ref{RDescr}) take the expressions: 
\bes 
\bar{\bar{t_I}} & = & {t_r} \left(1+{Q \over{R}}\right) 
\ \ \ \ \ \  , \ \ \ \ \ \ \label{RMAtch1}
\bar{t_{III}}  =  \left({1-{M\over{R}}}\right) {t_m}  \\
\bar{\bar{A_I}} & = & \bar{A_{II}} {\left({Q \over{R}}\right)}^Q 
{\left({t_r}\right)}^{R+Q} \ \ \ , \ \ \ 
A_{III}  =  \bar{A_{II}} {\left({R \over{M}}\right)}^{M} {{t_m}}^{R-M} 
\nonumber 
\ees
The descriptive variables take the form:
\bes \label{Rtbar} 
\bar{\bar{t}} & = &
 \left(1 + {Q\over{R}}{{t-t_r}\over{t_I-t_r}} \right)t_r \ \ \ \ , \ \ \ \ 
\bar{t}  =  {M \over{R}} t + \left(1 - {M \over{R}}\right) t_m 
\ees
The current time is transformed in this description:
$\bar{t_0} =  {M \over{R}} t_0 + \left(1 - {M \over{R}}\right) t_m $.
Also, the beginning of inflationary stage: $\bar{\bar{t_i}} =
 \left(1 + {Q\over{R}}{{t_i-t_r}\over{t_I-t_r}} \right)t_r $.
For usual values of temporal dependences, the matter dominated 
stage suffers a dilatation in these descriptive
variables $\left.{\Delta \bar{t}}\right|_{MAT}  =  \frac{M}{R} 
\left.{\Delta t}\right|_{MAT}$. For the inflationary stage, the
duration is modified as:
$\left.{\Delta {\bar{\bar{t}}}}\right|_{INF}  =  \frac{Q}{R}
\frac{t_r}{t_I-t_r} \left.{\Delta t}\right|_{INF}$. Dilatation
takes place also in this stage for $\frac{t_I}{t_r}<\frac{Q}{R}+1$.

\end{document}